\documentclass[11pt]{article}
\usepackage{graphicx}
\usepackage{amsmath}
\usepackage{amsxtra}
\newcommand{\reff}[1]{(\ref{#1})}
\newcommand{\beq}{\begin{equation}}
\newcommand{\eeq}[1]{\label{#1}\end{equation}}
\newcommand{\beg}{\begin{equation*}}
\newcommand{\eeg}{\end{equation*}}
\newcommand{\newd}{\textmd{d}}

\newcommand{\sumprime}{\sideset{}{'}\sum}

\newcommand{\bsplit}{\begin{split}}
\newcommand{\esplit}{\end{split}}

\begin{document}
\title{Casimir forces in Bose-Einstein condensates: finite size effects in three-dimensional rectangular cavities}
\author{\thanks{Email: aedery@ubishops.ca} Ariel Edery\\Bishop's University, Physics
Department\\Lennoxville, Quebec\\J1M1Z7}
\date{}
\maketitle
\begin{abstract}

\noindent The Casimir force due to {\it thermal} fluctuations (or
pseudo-Casimir force) was previously calculated for the perfect Bose
gas in the slab geometry for various boundary conditions. The
Casimir pressure due to {\it quantum} fluctuations in a
weakly-interacting dilute Bose-Einstein condensate (BEC) confined to
a parallel plate geometry was recently calculated for Dirichlet
boundary conditions. In this paper we calculate the Casimir energy
and pressure due to quantum fluctuations in a zero-temperature
homogeneous weakly-interacting dilute BEC confined to a parallel
plate geometry with periodic boundary conditions and include
higher-order corrections which we refer to as Bogoliubov
corrections. The leading order term is identified as the Casimir
energy of a massless scalar field moving with wave velocity equal to
the speed of sound in the BEC. We then obtain the leading order
Casimir pressure in a general three-dimensional rectangular cavity
of arbitrary lengths and obtain the finite-size correction to the
parallel plate scenario.
\end{abstract}

\newpage

\section{Introduction}

After nearly $50$ years since its prediction in 1948 by Casimir
\cite{Casimir}, the Casimir force has now been successfully measured
by a modern series of experiments starting with Lamoreaux's 1997
landmark experiment \cite{Lamoreaux} with a torsion pendulum which
reduced errors dramatically compared to the early 1958 experiment by
Spaarnay \cite{Spaarnay}. The force was subsequently measured more
precisely in 1998 using an atomic force microscope \cite{Mohideen}
and the measurements agreed with theoretical predictions to within
1\% after finite conductivity, roughness and temperature corrections
were taken into account. The Casimir force in the more difficult
parallel plate geometry was then successfully measured to 15\%
accuracy \cite{Onofrio}. Thus the modern era of precise Casimir
measurements was born. All the measurements of the Casimir force to
date have been limited to the case of the electromagnetic field.
However, experiments may soon measure the Casimir force for a
massless scalar field via quantum fluctuations in Bose-Einstein
condensates (BEC). It has been noted (see \cite{Roberts2}) that the
quasiparticle vacuum in a zero-temperature dilute weakly-interacting
BEC should give rise to a measurable Casimir force. The fact that a
scalar field propagates at the speed of sound in a BEC medium in
contrast to the speed of light in Minkowski spacetime does not
change anything fundamental in relation to the Casimir energy. If
the speed of propagation is constant in a given medium, the Casimir
energy in units of this speed will be the same value regardless of
whether the medium is spacetime or a BEC. Moreover, a generally
covariant action analogous to what we see in General Relativity
exists for scalar fields propagating in a particular fluid. The
Lagrangian is similar to that of a massless Klein-Gordon field with
the Minkowski metric $\eta_{\mu\,\nu}$ of spacetime replaced by an
effective or acoustic metric $g_{\mu\,\nu}$ \cite{Visser}. Quoting
directly from \cite{Visser2}, ``at low momenta linearized
excitations of the phase of the condensate wavefunction obey a
(3+1)-dimensional d'Alembertian equation coupling to a
(3+1)-dimensional Lorentzian-signature `effective metric' that is
generic, and depends algebraically on the background field.".

The Casimir force due to {\it thermal} fluctuations in a perfect
Bose gas confined to a slab geometry was recently calculated for
Dirichlet, Neumann and periodic boundary conditions \cite{Martin}
(see also comment \cite{Gambassi} on the work of \cite{Martin}). The
Casimir pressure due to {\it quantum} fluctuations for a
weakly-interacting dilute Bose-Einstein condensate confined to a
parallel plate geometry was also recently calculated \cite{Roberts}.
In this paper, we extend the work of \cite{Roberts} on Casimir
forces in BEC's to include a general three-dimensional cavity and
consider periodic instead of Dirichlet boundary conditions. We first
calculate the Casimir pressure due to quantum fluctuations of the
quasiparticle vacuum in a zero-temperature homogeneous dilute
weakly-interacting BEC confined to a ``parallel plate" geometry with
periodic instead of Dirichlet boundary conditions. We show that the
leading order term for the Casimir energy is equal to that of a
massless scalar field moving with wave velocity equal to the speed
of sound in the BEC. We also obtain the much smaller `Bogoliubov'
corrections due to the nonlinearity of the Bogoliubov dispersion
relation. We then generalize the results to a three-dimensional
``rectangular" cavity of arbitrary lengths and obtain the leading
order finite size corrections to the parallel plate scenario for
periodic boundary conditions. The quotes around ``rectangular" or
``parallel plate" are simply to remind the reader that with periodic
boundary conditions we are of course dealing with a hypertoroidal
geometry. We drop the quotes from now on.

To maintain a homogeneous gas in a slab geometry requires periodic
boundary conditions. Though such boundary conditions constitute an
idealization, it enables one to obtain relevant analytical results
for the Casimir pressure of the BEC prior to a numerical analysis of
the Casimir force in a non-homogeneous gas confined to a disk-like
geometry via an anisotropic harmonic potential.

\section{Casimir pressure for BEC in parallel plate geometry and massless scalar
fields}

Consider a weakly interacting BEC characterized by an interparticle
contact pseudopotential, $8\,\pi\,a\,\delta^{(3)}(r)$, where $a$ is
the 2-particle positive scattering length (we work in units of
$\hbar=1$ and $2m=1$). For this delta function potential, the
contribution $E$ of the depletion to the ground state energy due to
quantum fluctuations in a zero-temperature untrapped homogeneous
dilute weakly-interacting BEC is given by \cite{Lee, Lifshitz} \beq
E = \dfrac{1}{2}\sum_{\bf{k}\ne 0} E(k) =
\dfrac{1}{2}\sum_{\bf{k}\ne 0} ( \,k\,\sqrt{k^2+2\,\mu}-k^2-\mu\,)
\eeq{ground} where $\mu\equiv 8\,\pi\,a\,\rho$, $N$ is the number of
atoms, $\rho$ is the density $N/V$ and \beq E(k)\equiv E_B -k^2-\mu
= k\,\sqrt{k^2+2\,\mu} -k^2-\mu \,. \eeq{EK} $E_B\equiv
k\,\sqrt{k^2+2\,\mu}$ is the Bogoliubov dispersion relation.

With periodic boundary conditions, the homogeneous gas is confined
to a parallel plate geometry with a trapping potential which is
zero everywhere. It is worth noting that for quantum fluctuations
to manifest themselves, the plate separation $\newd$ must be much
greater than the healing length i.e. $\newd\!\!>\!>\mu^{-1/2}$. In
an $L_1\times L_2\times
\newd$ hypertoroidal space, $k^2 = (2\,n\,\pi/\newd)^2 +
(2\,n_1\,\pi/L_1)^2 + (2\,n_2\,\pi/L_2)^2$ where $(n,n_1,n_2) \ne
(0,0,0)$ and $n, n_1$ and $n_2$ are integers that run from
$-\infty$ to $+\infty$.
 Both the volume
$V=L_1\,L_2\,\newd$ and the number of atoms $N$ are assumed large
with the density $\rho=N/V$ low enough that the dilute condition
$\sqrt{\rho\,a^3}\!<\!<\!1$ is satisfied. The triple sum for the
ground state energy \reff{ground} can be broken up in the
following convenient fashion:
\begin{eqnarray} \sum_{\bf{k}\ne 0}
E(k)&=&\sum_{n=-\infty}^{\infty}\sum_{n_1=-\infty}^{\infty}\sum_{n_2=-\infty}^{\infty}
E(k)+\mu \nonumber\\&=&
8\sum_{n=1}^{\infty}\sum_{n_1=0}^{\infty}\sum_{n_2=0}^{\infty}
E(k) + 8 \sum_{n_1=0}^{\infty}\sum_{n_2=0}^{\infty} E(k)+ 8\,\mu
\,. \label{sumbreak}
\end{eqnarray}

For parallel plates, $L_1$ and $L_2$ are very large (infinite
limit) and the sums over $n_1$ and $n_2$ become integrals. The
double sum in \reff{sumbreak} becomes a double integral and does
not contribute to the Casimir energy being purely a continuous
term. The constant $8\mu$ also does not contribute to the Casimir
energy. The relevant term for the Casimir energy, the triple sum
in \reff{sumbreak}, becomes a sum over a double integral and $k^2$
reduces to $(2\,n\,\pi/\newd)^2 +r^2$. The relevant energy $E$ per
unit area is then given by \beq
\dfrac{1}{L_1\,L_2}\,E=\sum_{n=1}^{\infty} \,f(n)\eeq{EE} where
\beq
\begin{split}
f(n) & \equiv \dfrac{1}{2\,\pi} \int_0^{\infty}
\!\Bigg(\Bigg[\Big(\dfrac{4\,n^2\,\pi^2}{\newd^2} \!+r^2\Big)^2
\!+ 2\,\mu\,\Big(\dfrac{4\,n^2\,\pi^2}{\newd^2}\!
+r^2\Big)\Bigg]^{1/2}\!\!\!\!\!-\dfrac{4\,n^2\,\pi^2}{\newd^2} \!
-\!r^2 \!-\mu\!\Bigg) r\,dr
\\& = \dfrac{\mu^2}{\pi}\int_{\tfrac{2\,n^2\,\pi^2}{\newd^2\,\mu}}^{\Lambda} \,\,\left(\sqrt{u^2 +u} -u -\dfrac{1}{2}\,\right)\,du \,.
\end{split}
\eeq{fn} The integral can be evaluated but this is not necessary
since only the derivatives of $f(n)$ will be needed to determine the
Casimir energy. The parameter $\Lambda$ is an ultraviolet cut-off
introduced because the energy $E$ given by \reff{ground} and hence
$f(n)$ are formally divergent. There is nothing physical about this
divergence. It simply reflects that the delta function contact
pseudopotential cannot be na\"ively extrapolated to very high
momentum where the wavelength is comparable or smaller than the
interparticle spacing. This simple cut-off regularization scheme
cannot be used to calculate the actual finite energy $E$ of the
depletion because the result is clearly cut-off dependent. One must
either extract the low energy physics from the formally infinite
expression by making use of effective field theory and dimensional
regularization \cite{Andersen} or use the finite expression for $E$
derived in \cite{Lee, Lifshitz} via a modified pseudopotential.
However, to evaluate the Casimir energy, it is perfectly fine to use
\reff{fn} and in fact this is what was done in \cite{Roberts}. The
reason is that the Casimir energy is the difference between two
energies -- the discrete and the continuum -- and this difference
turns out to be independent of the cut-off $\Lambda$. The Casimir
energy in the BEC picks out the long wavelength behavior near $k=0$
and is therefore oblivious to the ultraviolet cut-off. We will show
explicitly that the terms in the Casimir energy correspond to terms
in the series expansion of $E(k)$ about $k=0$.

We now evaluate the Casimir energy density $E_{casimir}$. This is
equal to the difference between the energy density of the discrete
and continuum modes (both bulk and surface terms). As in
\cite{Roberts}, we evaluate the Casimir energy via the
Euler-Maclaurin formula \cite{Arfken}: \beq
\begin{split}
&E_{casimir}=\sum_{n=1}^{\infty} \,f(n) - \int_0^{\infty} f(n) +
\dfrac{1}{2} \,f(0)\\& = -\sum_{p=1}^{\infty}
\dfrac{B_{2p}}{(2p)!}f^{2p-1}(0) =
-\dfrac{B_2}{2!}\,f^1(0)-\dfrac{B_4}{4!}\,f^3(0)
-\dfrac{B_6}{6!}\,f^5(0) +\cdots\\&=
-\dfrac{\pi^2}{90}\dfrac{(2\,\mu)^{1/2}}{\newd^3}+\dfrac{2\,\pi^4}{315\,(2\,\mu)^{1/2}\,
\newd^{\,5}} +  O \,(\mu^{-3/2}\,d^{-7})\\&
=-\dfrac{\pi^2}{90}\dfrac{v}{\newd^3}+\dfrac{2\,\pi^4}{315\,v\,\newd^{\,5}}
+  O \,(\mu^{-3/2}\,d^{-7})
\end{split}
\eeq{PPP} where $v= (2\,\mu)^{1/2}$ is the speed of sound in the BEC
and $f^{2p-1}(0)$ are odd derivatives evaluated at zero. The leading
order result, $E_{scalar}= -\pi^2\,v/(90\,\newd^3)=-0.10966\,v/d^3$,
is exactly equal to the Casimir energy density of a massless scalar
field obeying a linear dispersion relation and confined to parallel
plates with periodic boundary conditions \cite{Svaiter,Ariel}. The
massless scalar field propagates with wave velocity $v$ equal to the
speed of sound in the BEC instead of the speed of light. The next
term is a `Bogoliubov' correction that arises because the Bogoliubov
dispersion relation is nonlinear and contains higher powers of $k$
when expanded about $k=0$. The magnitude of the ratio of the
correction $E_{Bogo}=2\,\pi^4/(315\,v\,\newd^{\,5})$ to the leading
order result $E_{scalar}$ is much smaller than unity since the plate
separation $\newd$ is assumed to be much greater than the healing
length $\mu^{-1/2}$ i.e. \beq \dfrac{E_{Bogo}}{|E_{scalar|}}
=\dfrac{2.82}{\mu\,\newd^{2}} <\!<\!1 \,.\eeq{corr} Therefore, to a
very good approximation,  the Casimir energy of the BEC corresponds
to quantum fluctuations of a massless scalar field propagating at
the speed of sound (i.e. acoustic phonons). This correspondence is
not only significant conceptually but also computationally. Results
on the Casimir energy of massless scalar fields can be applied to
the BEC to obtain leading order terms. We apply this principle in
the next section to obtain finite-size corrections to the parallel
plate scenario.

It is worth noting that the terms in the Casimir energy correspond
to terms in the series expansion of $E(k)$ about $k=0$ i.e. \beq
E(k)=k\,\sqrt{k^2+2\,\mu} -k^2-\mu= - \dfrac{v^2}{2} + v\,k - k^2
+ \dfrac{k^3}{2\,v} -\dfrac{1}{8\,v^3}\,k^5 + O(k^7)\,.\eeq{EK2}
In other words, the Casimir energy picks out the long wavelength
behavior near $k\!=\!0$ term by term. The first term in the above
expansion is a constant and does not contribute to the Casimir
energy. The next term proportional to $k$ is the linear dispersion
relation of a massless scalar field and is responsible for the
leading order Casimir energy $E_{scalar}$. The next term,
proportional to $k^2$, does not contribute to the Casimir energy.
This can be seen from the fact that $k^2=(2\,n\pi/\newd)^2+r^2$
contains a term proportional to $n^2$ whose odd-derivatives
evaluated at zero are zero or alternatively, from zeta function
regularization we obtain $\zeta(-2)=0$ . The $k^3$ term is
responsible for the correction $E_{Bogo}$. Let us show this
explicitly. The function $f(n)$ for $k^3=
\big(4\,n^2\pi^2/\newd^2+r^2\big)^{3/2}$ is $
f(n)=1/(2\,\pi)\int_0^{\Lambda}\big(4\,n^2\,\pi^2/\newd^2+r^2\big)^{3/2}\,r\,dr$.
The odd-derivative terms $B_{2p}\,f^{2p-1}(0)/(2p)!$ are all zero
except $-B_6\,f^5(0)/6!=4\,\pi^4/(315\,\newd^5)$. Multiplying by
the $k^3$ coefficient $1/(2\,v)$ yields
$2\pi^4/(315\,v\,\newd^{\,5})$ which is equal to $E_{Bogo}$.

The Casimir pressure is readily obtained by taking the derivative of
the Casimir energy density\footnote{Note that the derivative of the
speed of sound $v$ with respect to $\newd$ (by definition keeping
the number of atoms $N$ constant) is not zero but equal to
$-v/(2\,\newd)$. If the derivative had been assumed to be zero, the
numerical factor of the leading term in \reff{Pressure} would have
been be $-6/180$ instead of $-7/180$.}: \beq P_{casimir} =
\Big(\dfrac{-\partial E_{casimir}}{\partial
\newd}\Big)_N  = -\dfrac{7\,\pi^2}{180}\dfrac{v}{\newd^4} +
\dfrac{\pi^4}{35\,v\,\newd^{\,6}} + O \,( v^{-3}\,d^{-8}) \,.
\eeq{Pressure} The leading order Casimir pressure is \beq
P_{scalar}= -\dfrac{7\,\pi^2}{180}\dfrac{v}{\newd^4}\,. \eeq{pscale}
It is negative (attractive) and inversely proportional to the fourth
power of the distance as in Casimir's original calculation for the
electromagnetic field \cite{Casimir}. Casimir obtained
$-\pi^2\,c/(240\,\newd^4)$ for the attractive force per unit area
between parallel conductors in vacuum. The Casimir pressure in the
BEC is considerably weaker than in the electromagnetic case because
the speed of sound $v$ is orders of magnitude smaller than the speed
of light $c$. Although Casimir's original calculation was performed
in 1948, one had to wait nearly 50 years before the Casimir force
between two conductors was conclusively confirmed by experiments
starting with Lamoreaux's 1997 landmark experiment with a torsion
pendulum \cite{Lamoreaux} and then by Mohideen et {\it al.} 1998
experiment with an atomic force microscope \cite{Mohideen}. Here
also, one can expect that theory is considerably ahead of
experiments and that measuring the much smaller Casimir force in a
BEC is something for the next generation of BEC experiments.

\section{Leading order Casimir pressure in three dimensional cavity and finite-size corrections}

In this section we obtain expressions for the leading order
Casimir pressure in a zero-temperature homogeneous dilute
weakly-interacting BEC confined to a three-dimensional rectangular
cavity of arbitrary dimensions $L_1\times L_2\times L_3$ with
periodic boundary conditions (all three lengths are assumed to be
much greater than the healing length). We then determine
finite-size corrections to the parallel plate result $E_{scalar}$
and $P_{scalar}$ obtained in the last section. As in the previous
section, the only parameter of the BEC that enters in our
calculations is the speed of sound $v= (2\,\mu)^{1/2}$. We take
advantage of the equality between the Casimir energy of a massless
scalar field and the leading order Casimir energy of a BEC by
making use of recently derived formulas for the Casimir energy of
massless scalar fields propagating with speed $v$ in a
$d$-dimensional rectangular cavity of arbitrary lengths
$L_1,L_2,\ldots,L_d$ \cite{Ariel}. The speed $v$ in this context
is the speed of sound in the BEC. The Casimir energy in
\cite{Ariel} was conveniently expressed as a compact analytical
part plus remainder. For periodic boundary conditions (inserting
$\hbar$) the Casimir energy is given by \cite{Ariel}: \beq
\begin{split}&E_{p_{ _{\,L_1,\ldots, L_d}}}\,(d) =-\hbar\,\pi\, v\sum_{j=0}^{d-1}\dfrac{L_1\ldots L_j}{(L
_{j+1})^{j+1}}\Big( \Gamma(\tfrac{j+2}{2})\,\pi^{\frac{-j-4}{2}}\,
\zeta(j+2) + R_j \Big)\,.
\end{split}\eeq{epfinal}
$R_j$ is a remainder term expressed as an exponentially fast
converging sum:\beq R_j
=\sum_{n=1}^{\infty}\,\sumprime_{\substack{l_i=-\infty\\i=1,\ldots,
j}}^{\infty}\dfrac{2\,(n\,L_{j+1})^{\tfrac{j+1}{2}}\,\,K_{\tfrac{j+1}{2}}
\big(\,\tfrac{2\pi\,n}{L_{j+1}}\sqrt{(\ell_1\,L_1)^2+\cdots+(\ell_j\,L_j)^2}\,\,\,\big)}
{\pi\,
\left[(\ell_1\,L_1)^2+\cdots+(\ell_j\,L_j)^2\right]^{\tfrac{j+1}{4}}}\,\,.
\eeq{rjbb} The prime means that the case where all $\ell$'s are
zero is excluded. There is no remainder for $j=0$ (it starts at
$j=1$). As discussed in \cite{Ariel}, the remainder term is small
if we label the longest length $L_1$, the next greatest length
$L_2$, etc. In three dimensions, we therefore label the lengths
such that $L_1\ge L_2\ge L_3$. There is a clear physical
interpretation to the analytical and remainder part in
\reff{epfinal} which is discussed in section 3 of \cite{Ariel}.
The analytical part is the sum of individual parallel plate
energies out of which the rectangle is constructed while the
remainder is a small contribution due to the nonlinearity of the
energy. For a cube with periodic boundary conditions the remainder
is $1.5\%$ of the Casimir energy (a table of numerical results can
be found in section 5 of \cite{Ariel}). For three arbitrary
lengths, the remainder is even smaller than in the case of the
cube.

From \reff{epfinal} we easily obtain the Casimir energy in three
dimensions ($d=3$) for periodic boundary conditions: \beq E_{p} =
\hbar\,v\,\Big(\,\dfrac{-\pi}{6\,L_1}-\dfrac{\zeta(3)}{2\,\pi}\dfrac{L_1}{L_2^2}-\dfrac{\pi^2}{90}\dfrac{L_1\,L_2}{L_3^3}
-\dfrac{\pi\,L_1}{L_2^2}\,R_1(L_1,L_2)
-\dfrac{\pi\,L_1\,L_2}{L_3^3}\,R_2(L_1,L_2,L_3)\Big)\,.
\eeq{epfinal2} The remainders $R_1$ and $R_2$ are sums over
modified Bessel functions given by \reff{rjbb} i.e.
\beq\begin{split}
&R_1(L_1,L_2)=\sum_{n=1}^{\infty}\sum_{\ell=1}^{\infty}\,\dfrac{4\,n}{\pi\,\ell}\,\dfrac{L_2}{L_1}\,K_1\Big(2\,\pi\,n\,\ell\,\dfrac{L_1}{L_2}\Big)\\&
R_2(L_1,L_2,L_3)=\sum_{n=1}^{\infty}\,\sumprime_{\ell_1,\ell_2=-\infty}^{\infty}\!\!
\dfrac{2\,n^{3/2}\,K_{3/2}\Big(2\,\pi\,n\,\sqrt{\Big(\dfrac{\ell_1\,L_1}{L_3}\Big)^2+\Big(\dfrac{\ell_2\,L_2}{L_3}\Big)^2}\,\,\Big)}
{\pi\left[\Big(\dfrac{\ell_1\,L_1}{L_3}\Big)^2+\Big(\dfrac{\ell_2\,L_2}{L_3}\Big)^2\right]^{3/4}}\,.
\end{split}\eeq{RRR}
Note that only the ratios of lengths appear in the remainders and
that $L_1/L_3\ge 1$, $L_2/L_3\ge 1$ and $L_1/L_2 \ge 1$ since the
lengths are labeled such that $L_1\ge L_2\ge L_3$. The Bessel
function $K_1\big(2\,\pi\,n\,\ell\,L_1/L_2\big)$ has a maximum value
of $9.87\times10^{-4}$ which occurs when $n\!=\!\ell\!=\!1$ and
$L_1/L_2=1$. As the ratio of lengths increases, the Bessel function
decreases exponentially fast and becomes tiny quickly. For example,
if $L_1/L_2=10$, then $K_1(2\,\pi\,10)$ is of order $10^{-28}$. One
obtains the same order of magnitude for $K_{3/2}(2\,\pi\,10)$.

In the parallel plate scenario, the plate separation is the
smallest of the three lengths and is therefore $L_3$. The Casimir
energy per unit area is then \beq\begin{split}&\dfrac{1}{L_1L_2}
\,E_p =
\dfrac{\hbar\,v}{L_3^3}\Big(\,-\dfrac{\pi^2}{90}-\dfrac{\zeta(3)}{2\,\pi}\Big(\dfrac{L_3}{L_2}\Big)^3
-\dfrac{\pi}{6}\Big(\dfrac{L_3}{L_1}\Big)^2\dfrac{L_3}{L_2} +R
\Big)
\end{split}\eeq{Epdensity}
where $R$ is the remainder contribution given by \beq R=-\pi
\Big(\dfrac{L_3}{L_2}\Big)^3\,R_1(L_1,L_2)
-\pi\,R_2(L_1,L_2,L_3)\,.\eeq{RR} We recognize the leading term as
the parallel plate result\\
$E_{scalar}=-\pi^2\,\hbar\,v/(90\,\newd^3)$ with $\newd=L_3$. The
other three terms are the finite size corrections to the Casimir
energy. Let us now calculate the pressure along the plate
separation $L_3$. This is given by the negative derivative of the
Casimir density with respect to $L_3$. Keep in mind that the
velocity $v$ has a dependence on $L_3$ and that $\partial
\,v/\partial\,L_3 = -v/(2L_3)$. The Casimir pressure on the plates
is given by : \beq\begin{split}
P&=-\dfrac{1}{L_1\,L_2}\dfrac{\partial\,E_p}{\partial\,L_3}
\\&=\dfrac{\hbar\,v}{L_3^4}\Big(-\dfrac{7\,\pi^2}{180}-\dfrac{\zeta(3)}{4\,\pi}\Big(\dfrac{L_3}{L_2}\Big)^3
-\dfrac{\pi}{12}\Big(\dfrac{L_3}{L_1}\Big)^2\Big(\dfrac{L_3}{L_2}\Big)+P_{R}\Big)\,.
\end{split}\eeq{pfinite} $P_{R}$ is the contribution of the
remainder given by \beq
\begin{split}P_{R}=-\dfrac{\pi}{2}\Big(\dfrac{L_3}{L_2}\Big)^3 \, R_1-\dfrac{7\pi}{2}\,R_2 + \pi\,L_3\,R_2'
\end{split}\eeq{PR}
where the prime is a derivative with respect to $L_3$. These
derivatives are trivial to evaluate using the expressions
\reff{RRR} and yield again Bessel functions.

The interpretation of \reff{pfinite} is straightforward. When
$L_1\to \infty$ and $L_2\to \infty$, only the first term survives
and one recovers the leading order parallel plate result
$P_{scalar}=-7\,\pi^2\,\hbar\,v/(180\,\newd^4)$ with $\newd=L_3$.
The other three terms in \reff{pfinite} are the finite size
corrections and depend on the ratios $L_3/L_1$, $L_3/L_2$ and their
inverse. $P_R$ is orders of magnitude smaller than the other two
correction terms because, as we have seen, the Bessel functions and
their derivatives are tiny even in the case where the lengths $L_1$
and $L_2$ are equal to $L_3$. It is therefore an excellent first
approximation to drop $P_R$ to determine the finite size
corrections. Let us therefore make a few quick numerical
calculations. We quote results in units of $\hbar\,v/L_3^4$. In
these units $P_{scalar}=-7\,\pi^2/180=-0.383818$. Let us begin with
the extreme case where $L_1$ and $L_2$ are equal to $L_3$ i.e. a
``cube". Then $P= -7\,\pi^2/180-\zeta(3)/(4\,\pi) -\pi/12=
-0.741274$. The force remains attractive but is much stronger. It
constitutes a $93\%$ difference from $P_{scalar}$. Clearly, forces
depend on the finite sizes of $L_1$ and $L_2$. We are however
interested in finite-size corrections to the parallel plate scenario
where $L_1$ and $L_2$ are at least a few times longer than $L_3$.
Consider then the case where $L_1$ and $L_2$ are $5$ times longer
than $L_3$. Then $P=-0.386678$. This constitutes only a $0.75\%$
change from $P_{scalar}$. If $L_1$ and $L_2$ are $10$ times longer
than $L_3$, the pressure reduces to $P=-0.384175$, which constitutes
only a $0.09\%$ change from $P_{scalar}$. Therefore, for any
realistic parallel plate scenario where $L_1$ and $L_2$ are at least
a few times longer than $L_3$, the Casimir pressure is dominated by
$P_{scalar}$.

The result for the pressure given by \reff{pfinite}, including the
remainders, is a leading order result. It does not include the next
order, the Bogoliubov corrections which arise from the $k^3$ term in
the series expansion of $E(k)$ given by \reff{EK2}. We saw last
section that these Bogoliubov corrections are small. Nonetheless, if
we wish to include them, one needs to obtain the analog formulas to
\reff{epfinal} for higher power dispersion relations and this is
work for the future. However, we can already include the dominant
Bogoliubov correction. There is a separate Bogoliubov correction
associated with each of the four terms in \reff{pfinite}. If we
label the four terms $P_1, P_2, P_3$ and $P_4$ then there exists a
Bogoliubov correction for each term which we label
$P_{1_{Bogo}},P_{2_{Bogo}},P_{3_{Bogo}}$ and $P_{4_{Bogo}}$. The
dominant Bogoliubov correction, $P_{1_{Bogo}}$, is associated with
the leading term $P_1= -7\,\pi^2\,\hbar\,v/(180 \,L_3^4)$. Its value
has already been calculated in \reff{Pressure} i.e.
$P_{1_{Bogo}}=\pi^4/(35\,v\,\newd^{\,6})$. To match the units and
notation of this section, we insert $\hbar$ and $m$ and let
$\newd=L_3$. This yields
$P_{1_{Bogo}}=\hbar^3\,\pi^4/(140\,m^2\,v\,L_3^{\,6})$. So we can
already include the most important Bogoliubov correction into the
result \reff{pfinite} by replacing $P$ with $P + P_{1_{Bogo}}$.
$P_{1_{Bogo}}$ is clearly not a significant correction to $P_1$, but
it can still compete with the finite-size corrections $P_2$ and
$P_3$ when $L_1$ and $L_2$ are sufficiently larger than $L_3$. Of
course, both corrections are then small. To summarize, in the
parallel plate scenario $P_1=P_{scalar}$ is the dominant Casimir
pressure in the BEC and corresponds to the parallel plate result for
a massless scalar field. There are then small finite-size
corrections to $P_{scalar}$. If one is interested in keeping track
of tiny corrections, Bogoliubov corrections become comparable to
finite-size corrections when $L_1$ and $L_2$ are sufficiently larger
than $L_3$.

\subsection*{\center{Acknowledgments}} I wish to thank the Natural
Sciences and Engineering Council of Canada (NSERC) for their financial
support of this project.

\end{document}